\documentclass[aps,prd,twocolumn,superscriptaddress,showpacs,amsmath,amssymb,nofootinbib]{revtex4}

\usepackage{graphicx}
\usepackage{dcolumn}
\usepackage{bm}

\begin{document}

\title{A New Family of Light Beams and Mirror Shapes for Future  
LIGO Interferometers}

\author{Mihai Bondarescu}
 \email{mihai@tapir.caltech.edu}
  \homepage{http://www.tapir.caltech.edu/~mihai}
 \affiliation{High Energy Physics, California Institute of Technology, Pasadena, CA 91125}

\author{Kip S.\ Thorne}
 \affiliation{Theoretical Astrophysics, California Institute of Technology, Pasadena, CA 91125}

\date{\today}

\begin{abstract}
      
Advanced LIGO's present baseline design uses arm cavities with
Gaussian light beams supported by spherical mirrors.  Because Gaussian
beams have large intensity gradients in regions of high intensity, they
average poorly over  fluctuating bumps and valleys on the mirror surfaces, caused
by random thermal fluctuations ({\em thermoelastic noise}).  Flat-topped
light beams ({\em mesa beams}) are being considered as an alternative
because they average over the thermoelastic fluctuations much more effectively.
However, the proposed mesa beams are supported by nearly
flat mirrors, which experience a very serious {\em tilt instability}.   In this paper
we propose an alternative configuration in which mesa-shaped beams are supported by
nearly concentric
spheres, which experience only a weak tilt instability.  The tilt instability
is analyzed for these mirrors in a companion paper by Savov and Vyatchanin.
We also propose a one-parameter family of  light beams and mirrors in which,
as the parameter $\alpha$ varies continuously from 0 to $\pi$, the beams and supporting
mirrors get deformed
continuously from the nearly flat-mirrored mesa configuration (``FM'') at $\alpha=0$, to the
nearly concentric-mirrored mesa configuration (``CM'') at $\alpha=\pi$.  The FM and
CM configurations at the endpoints are
close to optically unstable, and as $\alpha$ moves away from 0 or $\pi$, the optical
stability improves. 

\end{abstract}

\pacs{Valid PACS appear here}

\maketitle

\section{\label{sec:level1} Introduction}

The initial gravitational-wave detectors in the  
Laser Interferometric Gravitational wave Observatory (LIGO) are now near
design sensitivity and are taking science data
\cite{LIGO}. The interferometers will be upgraded to a much more sensitive
{\it advanced-LIGO} design beginning in about 2007.  
Until 2003, the baseline
design for advanced LIGO used 
nearly flat but spherical mirrors in its arm cavities.   However, in 2003, Sidles and
Sigg \cite{Sidles} showed that these mirrors experience a strong tilt
instability: when the mirrors are tilted symmetrically, the light beam slides across
their surfaces to an off-center location and its light pressure then pushes hard to increase
the tilt.  Sidles and Sigg proposed switching to mirrors that are segments of nearly
concentric spheres (radii of curvature slightly larger than half the cavity length);
such mirrors, they showed, can support Gaussian beams of the same (large) radius
as the baseline design, while experiencing a much weakened tilt instability.  This triggered
a change of the baseline design for advanced LIGO to nearly concentric, spherical mirrors.

Gaussian beams have the serious disadvantage that, because of their steep
intensity gradient over most of the beam's area, they average poorly over the
fluctuating bumps and valleys on the mirrors' surfaces that are caused by thermal
fluctuations ({\em thermoelastic noise}).  
O'Shaughnessy and Thorne \cite{OT1,OT2,DOSTV} have proposed improving the averaging
and thereby reducing the thermoelastic noise substantially, by replacing 
the arm cavities' Gaussian beams by flat-topped beams ({\em mesa beams}, as Willems
has named them), which 
are supported by nearly flat, Mexican-hat-shaped mirrors.
O'Shaughnessy, Strigin and Vyatchanin \cite{Richard,DOSTV} have shown that
the thermoelastic noise for mesa beams is three times weaker in noise power
than for the baseline Gaussian beams (with beam sizes that produce the same
diffraction losses), and correspondingly that the mesa-beam interferometers could see
farther into the universe, producing event rates for inspiraling binaries about three
times higher than the baseline Gaussian-beam design.  D'Ambrosio et.\ al.\ 
\cite{Erika,Richard,DOSTV} have shown that the nearly flat mesa-beamed mirrors
are practical in all respects that could be analyzed theoretically, and D'Ambrosio,
DeSalvo, Simoni and Willems \cite{Willems} are building a prototype optical cavity with mesa beams,
to explore practical issues experimentally.

Unfortunately, the nearly  
flat, Mexican-hat-shaped mirrors (``FM'') proposed by O'Shaugnessy and Thorne  
to support mesa beams (Sec.\ \ref{sec:FM} below), like the nearly flat, 
spherical mirrors of the 
pre-2003
baseline
design, experience a severe tilt instability (Vyatchanin \cite{Sergey}; 
Savov and Vyatchanin \cite{Vyatchanin}).  In this paper, motivated by the Sidles-Sigg
result that, for Gaussian beams and spherical mirrors, the tilt instability
is greatly weakened by switching from nearly flat to nearly concentric mirrors, we
propose (Sec.\ \ref{sec:CM}) a new, nearly concentric mirror design (``CM'') that 
supports mesa beams.  In a companion paper, Savov and Vyatchanin \cite{Vyatchanin}
show that the tilt instability is weaker for these CM mirrors than for any other mirrors
thus far considered --- FM, nearly-flat spherical, and nearly-concentric spherical.

\begin{figure}
\includegraphics[width=0.45\textwidth]{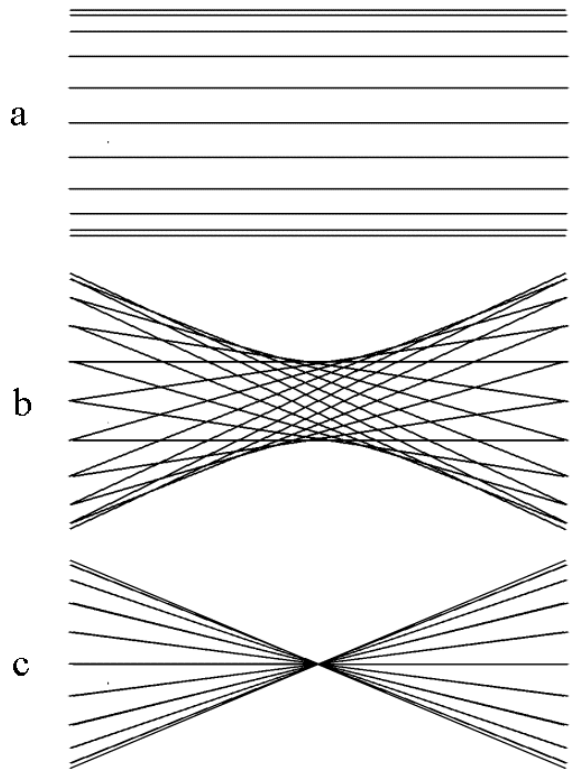}
\caption{Optical axes of the families of minimal Gaussians
beams used to construct: (a) an FM mesa beam \cite{Richard}, denoted in
this paper $\alpha=0$; (c) our
new  CM mesa beam, denoted $\alpha=\pi$;  (b) our new family of
hyperboloidal beams, which deform, as $\alpha$ varies from $0$ to $\pi$, from
a FM beam (a) into a CM beam (c).}
\label{fig:opticaxes}
\end{figure}

In Secs.\ \ref{sec:FM} and \ref{sec:CM}, we mathematically construct 
our FM and CM beams and their Mexican-hat mirror shapes  by superposing minimal-radius
Gaussian beams with optic axes that are the generators of cylinders (for FM) 
and of cones (for CM); Fig.\ \ref{fig:opticaxes}a,c.  In Sec.\ \ref{sec:Hyper} we
introduce a one-parameter family of ``hyperboloidal'' light beams and supporting mirrors,
computed by superposing minimal Gaussians whose optic axes are the generators
of hyperboloids; Fig.\ \ref{fig:opticaxes}b.  For each hyperboloidal beam, the 
hyperboloid's generators (minimal-Gaussian optic axes) have a fixed twist angle 
$\alpha$.  As $\alpha$  is varied
continually, the hyperboloidal light beams deform continually from mesa-shaped
FM form (at  $\alpha = 0$, where the hyperboloids are cylinders) to sharply peaked
Gaussian form (at $\alpha = \pi/2$), to mesa-shaped
CM form (at $\alpha = \pi$,
where the hyperboloids degenerate to cones).  

\section{Mesa beams supported by nearly flat mirrors (FM Beams; $\alpha=0$)}
\label{sec:FM}

The mesa beams supported by nearly flat Mexican-Hat mirrors (``FM'' beams) can be constructed 
mathematically by a procedure due to O'Shaughnessy and Thorne \cite{OT2,Richard}:  
One superposes minimal Gaussian beams\footnote{By ``minimal Gaussian beam'' we mean the fundamental, TEM00 mode of an optical resonator with spherical mirrors, with the mirror radii of curvature adjusted so the Gaussian-shaped intensity distributions on the two mirrors are identical and have the minimum possible radii $b$ at the $1/e$ point of the intensity distribution.}  with their optic axes all parallel to the cavity axis and
distributed uniformly
over a disk with some radius $D$, as shown in Fig.\ \ref{fig:opticaxes}a.  Each minimal Gaussian beam
(field) is given by
\begin{eqnarray}
\Psi(\varpi,\zeta) &=&\frac{\sqrt2}{\sqrt{1+ \zeta^2/\ell^2}} \exp\left\{ \frac{-\varpi^2/b^2}{1+\zeta^2/\ell^2}
\right. \label{Psi} \\
&& \left. + i \left[ \frac{\varpi^2/b^2}{\zeta/\ell + \ell/\zeta} - \arctan\left(\frac{\zeta}{\ell}\right) + \frac{2\ell\zeta}{b^2}\right]\right\}\:.
\nonumber
\end{eqnarray}
Here $\varpi$ is transverse distance from the beam's optic axis; 
$\zeta$ is distance parallel to the optic axis with $\zeta=0$ at the beam waist; $\ell \equiv L/2$
is half the length of LIGO's arm cavity (2 km) 
and is also equal to the beam's Rayleigh range; $b = \sqrt{\lambda L/2\pi}
= \sqrt{\lambda \ell /\pi} = 2.603$ cm (with $\lambda = 1.064 \mu$m the light wavelength)
is the radius, at the $1/e$ point of the beam's \emph{intensity} distribution, at the ends of the
cavity, i.e.\ at $\zeta = \ell$;  and $b$ is also the radius, at the $1/e$ point of the beam's \emph{amplitude} distribution, at the beam's waist, $\zeta = 0$.  
Note that the last phase factor, $2\ell \zeta/b^2$, is actually $k\zeta$ in disguise, with $k = 
2\pi/\lambda$ the light's wave number.  We adjust $\lambda$ or $\ell$ slightly so that at $\zeta = \ell$
and $\varpi=0$, $\Psi$ is real and positive, i.e.\ the sum of the last two phase factors is a multiple
of $2\pi$.  Then in the immediate vicinity of the mirror plane, at $\zeta = \ell + \delta\zeta$ (with $|\delta
\zeta| \ll b$), the minimal Gaussian has the simple form
\begin{equation}
\Psi(\varpi,\ell) = \exp \left[ \frac{-\varpi^2(1-i)}{2b^2} + ik\delta\zeta \right]\;,
\label{PsiEnd}
\end{equation}
with $k= 2\ell/b^2$.
(Here and throughout this paper we ignore fractional corrections of order $\lambda/b \sim b/\ell \sim 10^{-5}$.)
The mesa-beam (FM-beam) field, constructed by superposing minimal Gaussians as in Fig.\  \ref{fig:opticaxes}a, is
given by (Sec.\ IIA of \cite{DOSTV})
\begin{equation}
U_0(r,z,D) = \int_{\mathcal C_D} \Psi( \sqrt{(x-x_o)^2 + (y-y_o)^2}, z )\; dx_o dy_o\;.
\label{MesaFlat}
\end{equation}
Here $r = \sqrt{x^2+y^2}$ is radius from the cavity's central axis, the integral is over Cartesian coordinates $(x_o,y_o)$ of the Gaussians' optic axes, and
the integral extends over the interior of the disk $\mathcal C_D$ with radius $D$, i.e.\  $\sqrt{x_o^2 + y_o^2} \le D$.
The subscript $0$ on $U_0$ is the value $\alpha=0$ of the twist angle of the Gaussians' optic axes,
when one regards this FM beam from the viewpoint of the hyperboloidal family of beams (Sec.
\ref{sec:Hyper} below).

By inserting expression (\ref{PsiEnd}) with $\delta\zeta=0$ 
into Eq.\ (\ref{MesaFlat}), we obtain for the FM beam at the mirror plane $z = \ell$,
\begin{eqnarray}
&&U_0(r,\ell,D) = \label{MesaFlatEnd}\\
&&\quad \int_{\mathcal C_D} \exp \left[ \frac{-[(x-x_o)^2 + (y-y_o)^2][1-i]}{2b^2} \right] dx_o dy_o \nonumber\;.
\end{eqnarray}

\begin{figure}
\includegraphics[width=0.45\textwidth]{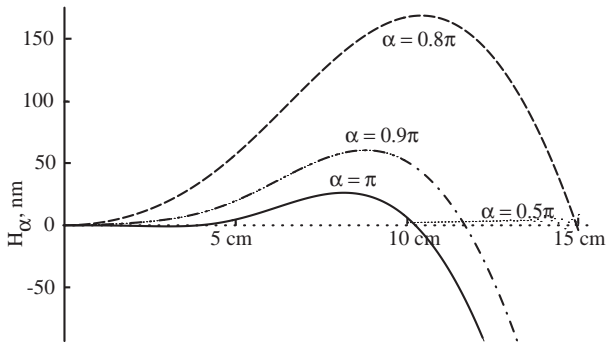}
\caption{The correction $H_\alpha(r)$ to the mirror shape for hyperboloidal beams with $D=10$ cm
and with twist angles $\alpha$ between $\pi/2$
and $\pi$.  For $\alpha = 0$, the correction is the negative of that for $\alpha=\pi$; for $\alpha=0.1 \pi$ 
it is the negative of that for $0.9\pi$; for any
$\alpha$ between 0 and $\pi/2$, it is the negative of that for $\pi-\alpha$.  For $\alpha=\pi$ (the
Mexican-hat correction for our new CM mesa beam), $H_0(r)$ drops to about $- 500$ nm (half
the wavelength of the light beam) at $r=16$ cm (the mirror's edge).
These corrections are added onto the
fiducial spheroidal shape $S_\alpha(r)$ [Eq.\ (\ref{Salpha})].
}
\label{fig:Shape}
\end{figure}

The mirror surface must coincide with a phase front of this mesa beam (FM beam),
i.e.\ it must have a shape $z = \ell + H_0(r) $ such that $\arg [ U_0(r,\ell + H_0,D)] = {\rm constant} 
= \arg[U_0(0,\ell,D)]$.  In the vicinity
of the mirror the phase of each minimal Gaussian varies as $k\delta\zeta = k \delta z$ [Eq.\ (\ref{PsiEnd})], so
$\arg[U_0(r,\ell + H_0,D)] = \arg[U_0(r,\ell,D)] + kH_0$, and the shape of the mirror surface must be 
\begin{eqnarray}
H_0(r) = k^{-1}\left\{ \arg[U_0(0,\ell,D) - \arg[U_0(r,\ell,D)]\right\}\;.
\label{HMesaFlat}
\end{eqnarray}
By carrying out
the integral (\ref{MesaFlatEnd}) analytically in one dimension then numerically in the other,
and then inserting into Eq.\ (\ref{HMesaFlat}), O'Shaughnessy and Thorne find the ``Mexican-hat''
mirror shape $H(r)$ shown as a solid line in Fig.\ \ref{fig:Shape}.
                                                                                    
\begin{figure}
\includegraphics[width=0.45\textwidth]{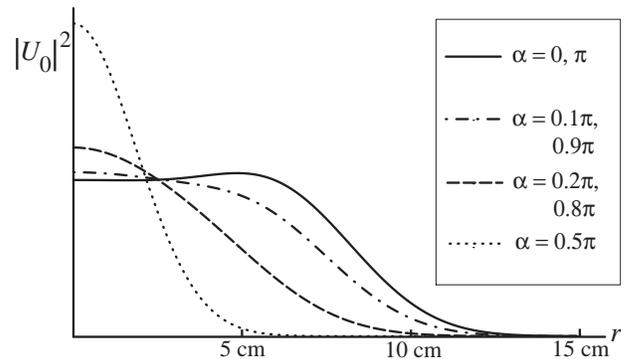}
\caption{The light beam's un-normalized intensity $|U_\alpha |^2$ 
as a function of radius $r$ on the mirror,  
for hyperboloidal beams with $D=10$ cm and various twist angles $\alpha$.
For $\alpha = 0$ and $\pi$, the intensity has the mesa shape; for $\alpha = 0.5$
it is a minimal Gaussian. }
\label{fig:Intensity}
\end{figure}

To high accuracy, the field $U_0$ on the mirror surface differs from that on the plane $z=\ell$ only
by the phase factor $e^{i k H_0(r)}$, so the intensity distribution on the mirror is the
same as at $z=\ell$; i.e., it is 
\begin{equation}
I_0 (r) \propto |U_0(r,\ell,D)|^2\;.
\label{I0}
\end{equation}
This intensity has the mesa shape shown as a solid line in  Fig.\ \ref{fig:Intensity}.

\section{Mesa beams supported by nearly concentric mirrors (CM Beams; $\alpha=\pi$)}
\label{sec:CM}

Our proposed mesa beams with nearly concentric, spherical mirrors (``CM'' beams) can
be constructed by overlapping minimal
Gaussians whose optic axes all pass through the center of the cavity [Fig.\ \ref{fig:opticaxes}(c)], 
and are distributed
uniformly inside a cone with angular radius $\Theta = D/\ell$

It should be clear from this construction that the resulting mesa beam will have a beam radius approximately equal to $D$ at the mirrors $(z=\ell)$, and approximately equal to $b$ (the minimal
Gaussian radius) at the
cavity's center, $z=0$.  For advanced LIGO, $D$ is approximately $4b$ \cite{DOSTV}, so the waist of
this mesa beam will be approximately 4 times narrower than the beam  on the mirrors.  This
contrasts with a mesa beam supported by nearly flat mirrors (previous section), for which
the waist is only slightly narrower than the beam on the mirrors.

Near the mirrors, the phase fronts of this CM beam will be nearly concentric spheres centered
on the point $(r,z) = (0,0)$ through which the Gaussians' optic axes pass, 
so we shall evaluate the CM field as a function of radius $r$ on this \emph{fiducial sphere},
\begin{equation}
z= S_\pi(r) \equiv \sqrt{\ell^2 - r^2} \simeq \ell - r^2/2\ell\;.
\label{sphere}
\end{equation}
[Here and below we use a subscript $\pi$ to denote mesa-beam quantities with nearly concentric
mirrors, i.e.\ CM quantities.  This is because the minimal Gaussians used to generate the CM
beam have twist angles $\alpha=\pi$; see Sec.\ \ref{sec:Hyper}.]
For each minimal Gaussian, this fiducial sphere bends away from the Gaussian's transverse plane
by an amount $\delta \zeta = -\varpi^2/2\ell$, so on this fiducial sphere the Gaussian's
phase factor $k\delta\zeta = (2\ell/b^2)\delta\zeta$ is equal to $- \varpi^2/b^2$.  As a result,
the Gaussian field on the fiducial sphere is [cf.\ Eq.\ (\ref{PsiEnd})]
\begin{eqnarray}
\Psi(\varpi, S_\pi) &=& \exp \left[ \frac{-\varpi^2(1-i)}{2b^2} +i k \delta\zeta \right] \nonumber\\
&=& \exp \left[ \frac{-\varpi^2(1+i)}{2b^2}\right]\;.
\label{PsiSphere}
\end{eqnarray}
Correspondingly, these minimal Gaussians superpose, on the fiducial sphere, to produce a
CM field given by
\begin{eqnarray}
&&U_\pi (r,S_\pi,D) = \label{MesaConcEnd}\\
&&\quad \int_{\mathcal C_D} \exp \left[ \frac{-[(x-x_o)^2 + (y-y_o)^2][1+i]}{2b^2} \right] dx_o dy_o \nonumber\;.
\end{eqnarray}
Notice that this $U_\pi (r,S_\pi,D)$ is the complex conjugate of the FM field $U_0(r,\ell,D)$ 
evaluated
on the transverse plane $z=\ell$ (the fiducial surface for the case of nearly flat mirrors); Eq.\ (\ref{MesaFlatEnd}).

As for the nearly flat (FM) case,  the phase of the CM field will vary, with distance $\delta z$ from the
fiducial sphere, nearly
proportionally to $k \delta z$; and correspondingly, the mirror's surface, $ \delta z = H_\pi(r)$ (a surface of constant
phase),  will be given by the analog of Eq.\ (\ref{HMesaFlat}):
\begin{equation}
H_\pi(r) = k^{-1}\left\{ \arg[U_\pi(0,S_\pi,D) - \arg[U_\pi(r,S_\pi,D)]\right\}\;.
\label{HMesaConc}
\end{equation}
Because $U_\pi(\varpi,S_\pi,D)$ is the complex conjugate of $U_0(\varpi,\ell,D)$, Eqs.\ (\ref{HMesaFlat}) and (\ref{HMesaConc}) imply that
\begin{equation}
H_\pi(r) = - H_0(r)\;.
\label{dualityH}
\end{equation}
In words:  to support mesa beams with the same beam size $D$ on their mirrors, the nearly concentric mirrors 
and the nearly flat mirrors must deviate from precisely concentric spheres $z= S_\pi(r) $ and precisely flat planes $z=\ell$ by equal and
opposite displacements $\delta z = H(r)$.   This fact was discovered in numerical work by one of 
us (MB) and was later proved numerically in a much wider context by Savov \cite{Vyatchanin}
and analytically by Agresti, d'Ambrosio, Chen and Savov \cite{Pavlin}, before we found the above demonstration. 

Because [to the accuracy of our analysis, $O(\lambda/b)$] the field $U_\pi$ is the same, aside from phase, on the mirror surface as on the fiducial
sphere $S_\pi$, the light's intensity distribution is the same on the mirror as on $S_\pi$:
\begin{equation}
I_\pi (r) \propto |U_\pi(r,S_\pi,D)|^2\;.
\label{Ipi}
\end{equation}
Moreover, because $U_\pi(r,S_\pi,D)$ is the complex conjugate of $U_0(r,\ell,D)$, they have the same
moduli and intensity distributions --- i.e., the CM beam has the same mesa-shaped intensity distribution
as the FM beam (solid curve in Fig.\ \ref{fig:Intensity} below).   This fact was discovered in numerical work by one of 
us (MB) and was later proved numerically in a much wider context by Savov \cite{Vyatchanin}
and analytically by Agresti, d'Ambrosio, Chen and Savov \cite{Pavlin}, before we found the above demonstration.

\section{Hyperboloidal beams supported by nearly spheroidal mirrors}
\label{sec:Hyper}

One can smoothly transform the FM beams into CM beams, and the 
in-between beams may be interesting for LIGO. In this section, we will focus on one way 
to make such a transformation. 

We will first look at a smooth deformation of the geometric body formed
by the optic axes of the minimal Gaussians that are used in constructing the
FM and CM beams. For a FM beam, the axes of the 
minimal Gaussians lie on coaxial cylinders, while for CM beams they 
lie on coaxial cones. It is well-known that one can smoothly deform a 
cylinder into a cone as follows. The generators of a cylinder of height 
$2\ell$ and radius $r$ (Fig.\ \ref{fig:opticaxes}a) are lines that join points with cylindrical
coordinates $(r,\phi,-\ell)$ on the 
base circle to points 
$(r,\phi,\ell)$ on the top circle. 
The generators of a symmetric cone of height $2\ell$ and end radii $r$ (Fig.\ \ref{fig:opticaxes}c) are 
lines that join points $(r,\phi,-\ell)$ and points $(r,\phi+\pi,\ell)$. A path from the cylinder to the
cone is given by a family of hyperboloids generated by lines that join 
points $(r,\phi,-\ell)$ and points $(r,\phi+\alpha,\ell)$ (Fig.\ \ref{fig:opticaxes}b). For $\alpha=0$ 
one obviously gets the cylinder and for $\alpha=\pi$, the cone.   

We therefore propose constructing a new two-parameter family of light beams, and the 
mirrors that support
these beams, using the O'Shaugnessy-Thorne technique of superposing minimal
Gaussians.  The parameters are $\{\alpha, D\}$, and for  given values of $\{\alpha,D\}$ the minimal
Gaussians have their optic axes uniformly distributed on the hyperboloid generators that reach
from  $(r,\phi,-\ell)$ to $(r,\phi+\alpha,\ell)$ (Fig.\ \ref{fig:opticaxes}c), with $\phi$
running from $0$ to $2\pi$ and $r$ confined to the interior of the disk $\mathcal C_D$, $r\le D$.  For
$\alpha=0$ these {\it hyperboloidal beams} will be mesa beams with nearly flat mirrors, i.e.\
FM beams.  For $\alpha=\pi$, they will be mesa beams with nearly concentric mirrors, i.e.\ CM
beams.  

We can construct explicit expressions for the shapes of the mirrors that support these hyperboloidal beams, and expressions
for the fields on those mirrors, using the same method as in the FM case (Sec.\  \ref{sec:FM})
and the CM case (Sec.\ \ref{sec:CM}):  Because the phase of each minimal Gaussian varies
nearly proportionally to $k \zeta$, the surface of constant phase at the mirror location will be 
nearly the same as the ``fiducial'' surface obtained by cutting off each Gaussian's optic axis at $\zeta = \ell$.
One can show that, with the optic axes being generators of hyperboloids, the surface formed by
their ends at constant distance $\zeta = \ell$ from the cavity's mid point is the fiducial spheroid
\begin{equation}
z = S_\alpha (r) \equiv  \sqrt{l^2-r^2 \sin^2(\alpha/2)} \simeq \ell - \frac{r^2 \sin^2(\alpha/2) }{2\ell}\;.
\label{Salpha}
\end{equation}

\begin{figure}
\includegraphics[width=0.45\textwidth]{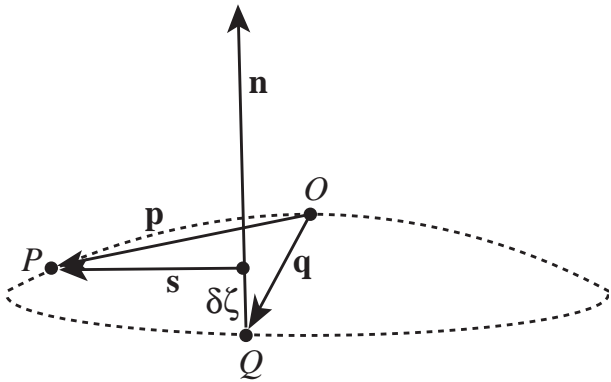}
\caption{Geometric construction for computing the hyperboloidal field $U_\alpha(r,S_\alpha,D)$ on
the fiducial spheroid $S_\alpha$ (a segment of which is shown dotted). }
\label{fig:Spheroid}
\end{figure}

We can compute our hyperboloidal field $U_\alpha(r,S_\alpha,D)$ on this fiducial spheroid by 
superposing our minimal Gaussians with the aid of Fig.\ \ref{fig:Spheroid}.  In this figure $P$ is the point
on the spheroid $S_\alpha$ at which we wish to compute the field.  The vector $\mathbf p$ reaching from
the spheroid's center point $O$ (the center of our hyperboloidal field's cross section) to $P$ has
Cartesian coordinates $\mathbf p = (r,0,\mathcal Z)$, where 
$\mathcal Z \equiv S_\alpha(r) - \ell = - (r^2/2\ell) \sin^2 (\alpha/2)$.  The optic axis of a minimal Gaussian, over which we will integrate, intersects 
$S_\alpha$ at the point $Q$, which has Cartesian coordinates $\mathbf q = (r_o \cos \phi_o, 
r_o \sin\phi_o, \mathcal Z_o)$, where 
$\mathcal Z_o = S_\alpha(r_o) - \ell =  - (r_o^2/2\ell) \sin^2 (\alpha/2)$.  The optic axis of this
minimal Gaussian points along the unit vector $\mathbf n = \{(r_o/2\ell)[\cos(\phi_o) - \cos(\phi_o-\alpha)],
(r_o/2\ell)[\sin(\phi_o) - \sin(\phi_o-\alpha)], 1\}$.  [Here as elsewhere we neglect corrections of order
$r/\ell \sim r_o/\ell \sim b/\ell \sim \lambda/b \sim 10^{-5}$.]  The vector $\mathbf s = \mathbf p - 
(\mathbf q +  \mathbf n \delta\zeta)$ reaches orthogonally from the minimal Gaussian's optic axis to
the point $P$.  The length of this vector is the radius $\varpi$ of $P$ as measured in cylindrical
coordinates centered on the minimal Gaussian's optic axis,
\begin{equation}
\varpi = |\mathbf s| \simeq |\mathbf p - \mathbf q| = \sqrt{r^2 + r_o^2 - 2 r r_o \cos\phi_o}\;,
\label{varpiHyperboloid}
\end{equation}
where the second expression, accurate to $O(\lambda/b)$, can be deduced from the above equations.
The distance $\delta\zeta$ along the optic axis $\mathbf n$, at which the normal $\mathbf s$ intersects
the axis, is determined by the orthogonality relation $\mathbf s \cdot \mathbf n = 0$:
\begin{equation}
\delta\zeta = \mathbf n \cdot (\mathbf p - \mathbf q) = {-1\over 2\ell} \left[ \varpi^2 \sin^2\left({\alpha\over2}
\right)+ r r_o \sin\alpha \sin\phi_o \right]\;.
\label{dzetaHyperboloid}
\end{equation}
The field $U_\alpha(r,S_\alpha,D)$ on the spheroid $S_\alpha$ is obtained by adding up the 
minimal Gaussians (\ref{PsiEnd}) with $\varpi$ and $\delta\zeta$ given by Eqs.\ (\ref{varpiHyperboloid})
and (\ref{dzetaHyperboloid}), and with $k = 2\ell/b^2$, and by then doing some simple algebra:
\begin{eqnarray}
&&U_\alpha(r,S_\alpha,D) = \int_0^D dr_o \int_0^{2\pi} d\phi_o  \exp\left[ i{rr_o\over b^2}
\sin\phi_o \sin\alpha \right.  \nonumber\\
&&\quad\quad\left.  -{(r^2 + r_o^2 -2r r_o \cos\phi_o)\over 2b^2} (1-i\cos\alpha) \right]\;.
\label{Ualpha}
\end{eqnarray} 
The radial integral can be carried out analytically yielding an expression involving error functions, and
the angular integral can then be done numerically.

The field (\ref{Ualpha}) cannot be sensitive to the chirality of the optic axes' twist, i.e.\ to the sign of $\alpha$, since it is a scalar complex function of $r$:  $U_{-\alpha} = U_\alpha$.  This tells us that
{\it the relevant range for $\alpha$ is $0$ to $\pi$}.  
Replacing $\alpha$ by $\pi - \alpha$ and changing the sign of $\alpha$ is equivalent to complex conjugating $U_\alpha$; therefore:
\begin{equation}
U_{-\alpha} = U_\alpha\;; \quad U_{\pi-\alpha} = U_\alpha^*\;.
\label{Usymmetries}
\end{equation}
For $\alpha=0$, the fiducial spheroid $S_0(r)$ is the transverse plane and the field (\ref{Ualpha}) is the FM mesa beam $U_0$ [Eq.\ (\ref{MesaFlatEnd})].  For $\alpha=\pi/2$, the fiducial spheroid $S_{\pi/2}$ is
a sphere of radius $R = L=2\ell$ (the distance between the mirrors), and both the radial and the angular integrals can be carried out analytically, giving for the field on that sphere
\begin{equation}
U_{\pi/2} = {\rm constant} \exp\left[{-r^2 /2b^2}\right]\;;
\label{Upitwo}
\end{equation}
this is precisely the minimal Gaussian beam [Eq.\ (\ref{PsiEnd}) with $\varpi=r$, evaluated at $k \delta \zeta = 
-k r^2/2 R = -(2\ell/b^2)r^2/4\ell = -r^2/2b^2$].  For $\alpha=\pi$, the fiducial spheroid $S_\pi(r)$ is 
a sphere with radius $\ell = L/2$ and the field (\ref{Ualpha}) is the CM mesa beam [Eq.\ (\ref{MesaConcEnd})].  Thus,
as $\alpha$ varies from $0$ to $\pi$, $U_{\alpha}$ deforms continuously from the FM mesa beam
$\alpha=0$,
through a set of hyperboloidal beams to a minimal Gaussian at $\alpha=\pi/2$, and on through
another set of hyperboloidal beams to the CM mesa beam $\alpha=\pi$.

As for the FM and CM beams, so also for the hyperboloidal beam (\ref{Ualpha}) (and for the same reasons), the
mirror's surface must be displaced longitudinally from the fiducial spheroid $z = S_\alpha(r)$ by
$\delta z = H_\alpha(r)$, where
\begin{equation}
H_\alpha(r) = k^{-1}\left\{ \arg[U_\alpha(0,S_\alpha,D) - \arg[U_\alpha(r,S_\alpha,D)]\right\}\;.
\label{Hhyperboloid}
\end{equation}
This equation and $U_{\pi-\alpha} = - U_\alpha^*$ [Eq.\ (\ref{Usymmetries})] tell us that
\begin{equation}
H_{\pi - \alpha} (r) = - H_{\alpha} (r)\;.
\label{Hsign}
\end{equation}
This is a special case of a duality relation discovered numerically by Savov and Vyatchanin
\cite{Vyatchanin} and proved analytically by Agresti, d'Ambrosio, Chen and Savov \cite{Pavlin}.
Figure \ref{fig:Shape} shows these mirror shape corrections for various $\alpha$'s.   
The light intensity on the mirrors is given by the obvious analog of Eqs.\ (\ref{I0}) and (\ref{Ipi}):
\begin{equation}
I_\alpha(r) \propto |U_\alpha(r,S_\alpha,D)|^2\;.
\label{Ialpha}
\end{equation}
This equation and $U_{\pi-\alpha} = - U_\alpha^*$ [Eq.\ (\ref{Usymmetries})] tell us that
\begin{equation}
I_{\pi-\alpha}(r) = I_\alpha(r)\;.
\label{Iinv}
\end{equation}
This is another special case of the duality relation discovered by Savov and Vyatchanin 
\cite{Vyatchanin}, and proved analytically by Agresti,   d'Ambrosio, Chen and Savov
\cite{Pavlin}.
The intensity distribution (\ref{Iinv}), (\ref{Ualpha}) is shown, for various $\alpha$, in Fig.\ 
\ref{fig:Intensity}.

\section{Conclusions}

For twist angles $\alpha$ near $0$ and $\pi$, the hyperboloidal beams introduced in 
this paper have the flat-top form needed to reduce thermoelastic noise in LIGO.  The radius
of the flat top is largest for 
$\alpha=0$ and $\alpha=\pi$ (the FM and CM mesa beams) and smallest for $\alpha=\pi/2$ (the minimal 
Gaussian).   

Because the mirrors are most nearly flat or concentric for the mesa configurations,
$\alpha = 0$ or $\pi$, those configurations are most nearly optically unstable.  (Near instability
goes hand in hand with large beams on the mirrors, which are needed to control thermoelastic noise.)

The results of  Savov and Vyatchanin \cite{Pavlin} suggest that the tilt instability is smallest for  
$\alpha=1$ and worst for $\alpha=\pi$.  

These considerations suggest that the optimal configuration for advanced LIGO will be near $\alpha
= \pi$, but whether the optimum is precisely at $\alpha=\pi$ (the CM configuration) or at some modestly
smaller $\alpha$ will depend on practical and thermoelastic-noise considerations not examined
in this paper.

\section{Acknowledgements}

We thank Pavlin Savov, Juri Agresti, Erika D'Ambrosio, Yanbei Chen, Geoffery 
Lovelace, and Poghos Kazarian for useful discussions 
and advice.  MB thanks Manuela Campanelli and the University of Texas at Brownsville, and Ed Seidel, Gabrielle Allen and the Center for Computation and Technology at Louisiana State University for 
helpful discussions and 
travel
support during this research.  This research was supported in part by NSF Grant  PHY-0099568.

\end{document}